\documentclass[useAMS,usenatbib]{mn2e}
\usepackage{graphicx}
\usepackage[usenames,dvipsnames]{color}
\def\hl{}

\title[KOI-13: spin-orbit resonance, TDV and secular perturbations]
{Spin--orbit resonance, transit duration variation and possible secular perturbations in KOI-13}
 
\author[Gy. M. Szab\'o et al.]{
Gy. M. Szab\'o$^{1,2}$\thanks{E-mail: szgy@konkoly.hu},
A. P\'al $^{1,3}$,
A. Derekas$^{1}$,
A. E. Simon$^{1,2}$,
T. Szalai$^{4}$,
L. L. Kiss$^{1,5}$\\
$^1$Konkoly Observatory of the Hungarian Academy of Sciences, PO. Box 67, H-1525 Budapest, Hungary\\
$^2$Department of Experimental Physics and Astronomical Observatory, University of Szeged, H-6720 Szeged, Hungary\\
$^3$Department of Astronomy, E\"otv\"os University, P\'azm\'any P\'eter s\'et\'any 1/A, 1117 Budapest, Hungary\\
$^4$Department of Optics and Quantum Electronics, University of Szeged, D\'om t\'er 9., H-6720 Szeged, Hungary\\
$^5$Sydney Institute for Astronomy, School of Physics A28, University of Sydney, NSW 2006, Australia
}

\begin{document}

\date{Accepted Received; in original form}

\pagerange{\pageref{firstpage}--\pageref{lastpage}} \pubyear{2010}

\maketitle

\label{firstpage}

\begin{abstract}

KOI-13 is the first known transiting system exhibiting light curve 
distortions due to gravity darkening of the rapidly rotating host star. 
In this paper we analyse publicly available $Kepler$ Q2--Q3 short-cadence
observations, revealing a continuous light variation 
with a period of $P_{\rm rot}=25.43\pm0.05$ hour and a half-amplitude of 
21\,ppm, which is linked to stellar rotation. This period is in exact 
5:3 resonance with the orbit of KOI-13.01, which is the first detection 
of a spin-orbit resonance in a host of a substellar companion. The 
stellar rotation leads to stellar oblateness, which is expected to cause 
secular variations in the orbital elements. We indeed detect the 
gradual increment of the transit duration with a rate of 
$(1.14\pm0.30)\times 10^{-6}$\,day/cycle. The confidence of this 
trend is 3.85-$\sigma$, the two-sided false alarm probability is 
0.012\%. We suggest that the reason for this variation is the expected 
change of the impact parameter, with a rate of 
${\rm d}b/{\rm d}t=-0.016 \pm 0.004/{\rm yr}$. Assuming $b\approx0.25$, KOI-13.01 may become a 
non-transiting object in $75-100$ years. The observed rate is compatible 
with the expected secular perturbations due to the stellar oblateness
yielded by the fast rotation.

\end{abstract}

\begin{keywords}
planetary systems
\end{keywords}

\section{Introduction}

KOI-13 (KIC 009941662) is a unique astrophysical laboratory of close-in companions in an 
oblique orbital geometry. The system consists of a widely separated 
common proper motion binary of A-type stars, one hosting a highly 
irradiated planet candidate with $P_{\rm orb}\approx 1.7626$ day 
(Borucki 2011). 
The host 
star of KOI-13.01 is the brighter component, KOI-13\,A, and rotates rapidly
($v\sin i \approx\,65 - 70\,{\rm km/s}$, Szab\'o et al. 2011). 
The transit curves 
show significant distortion that is stable in shape, and it 
is consistent with a companion orbiting a rapidly rotating star exhibiting gravity darkening by 
rotation (Barnes 2009).
\hl{Barnes et al. 2011 derived a projected alignment of $\lambda=23^\circ\pm4^\circ$ and the star's north pole is tilted away from the observer by $\psi=48^\circ\pm4^\circ$, therefore the stellar inclination, $i_*=43^\circ\pm4^\circ$ (assuming $M_*=2.05~\mathrm{M_{\sun}}$). The mutual inclination is $\varphi=54$--$56^\circ$. 
Companion is determined to have a mass of $9.2\pm1.1$~M$_J$ (Shporer et al. 2011) and between $4\pm 2$--$6\pm 3$~M$_J$ (Mazeh et al. 2011), and a radius of $1.44$~R$_J$ (Barnes et al. 2011).} On 23 September, 2011, new \emph{Kepler} photometry (Short Cadence Q3 data) became available. \hl{Here we suggest that the photometric data reveal the stellar rotation, and give observational evidence for transit duration variations (TDV) which are a sign of secular perturbations.}

\section{Rotation of the host star}

\begin{figure}
\includegraphics[bb=105 220 458 509,width=8.1cm,clip]{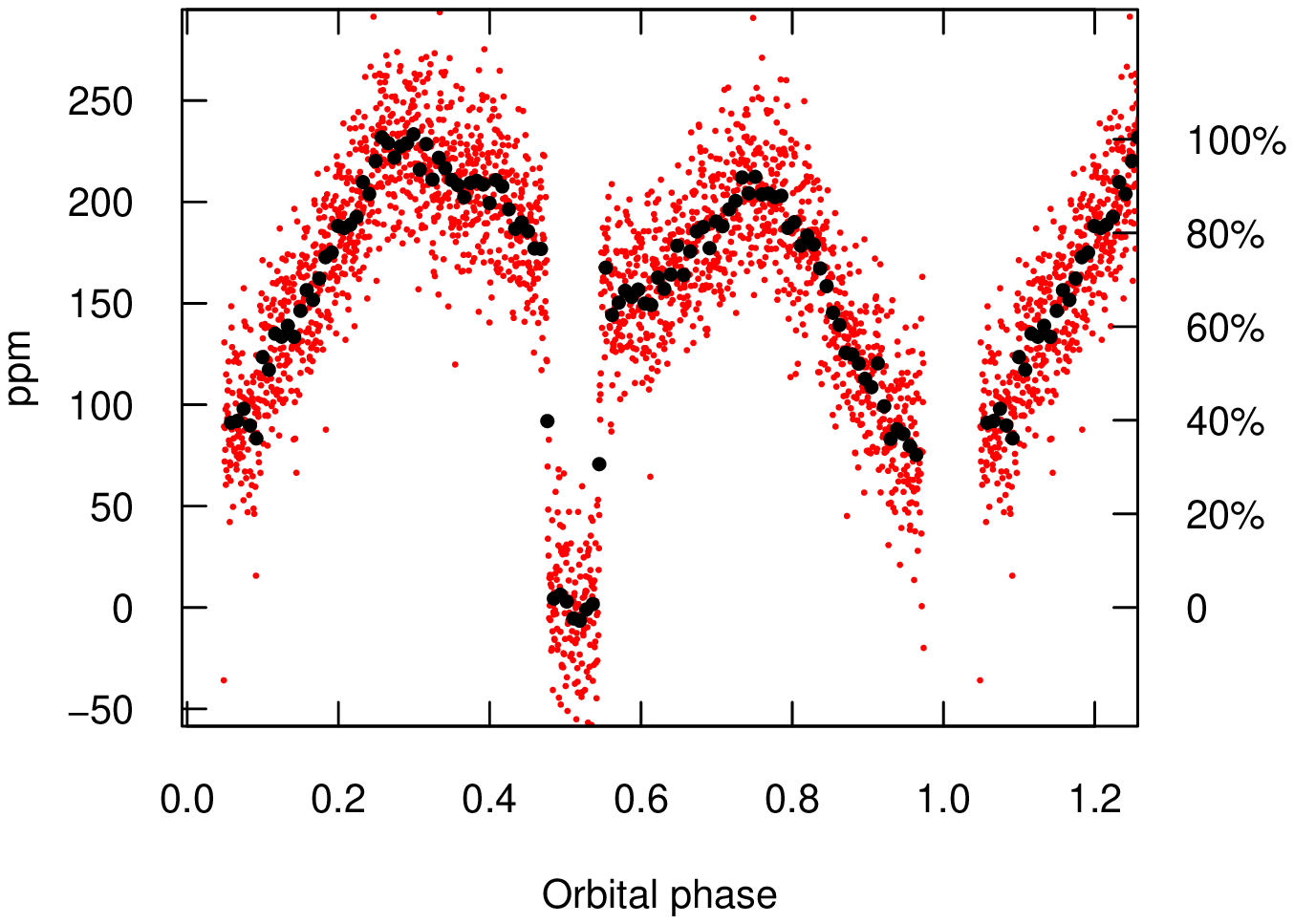}
\includegraphics[bb=108 292 458 450,width=8.1cm,clip]{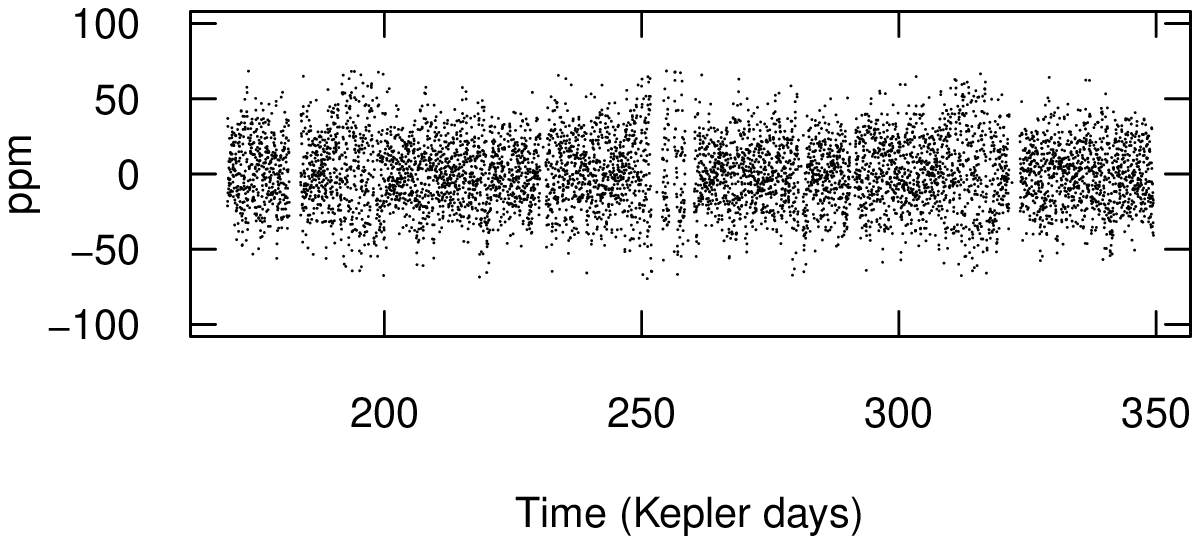}
\includegraphics[bb=70 240 466 457,clip,width=8.1cm]{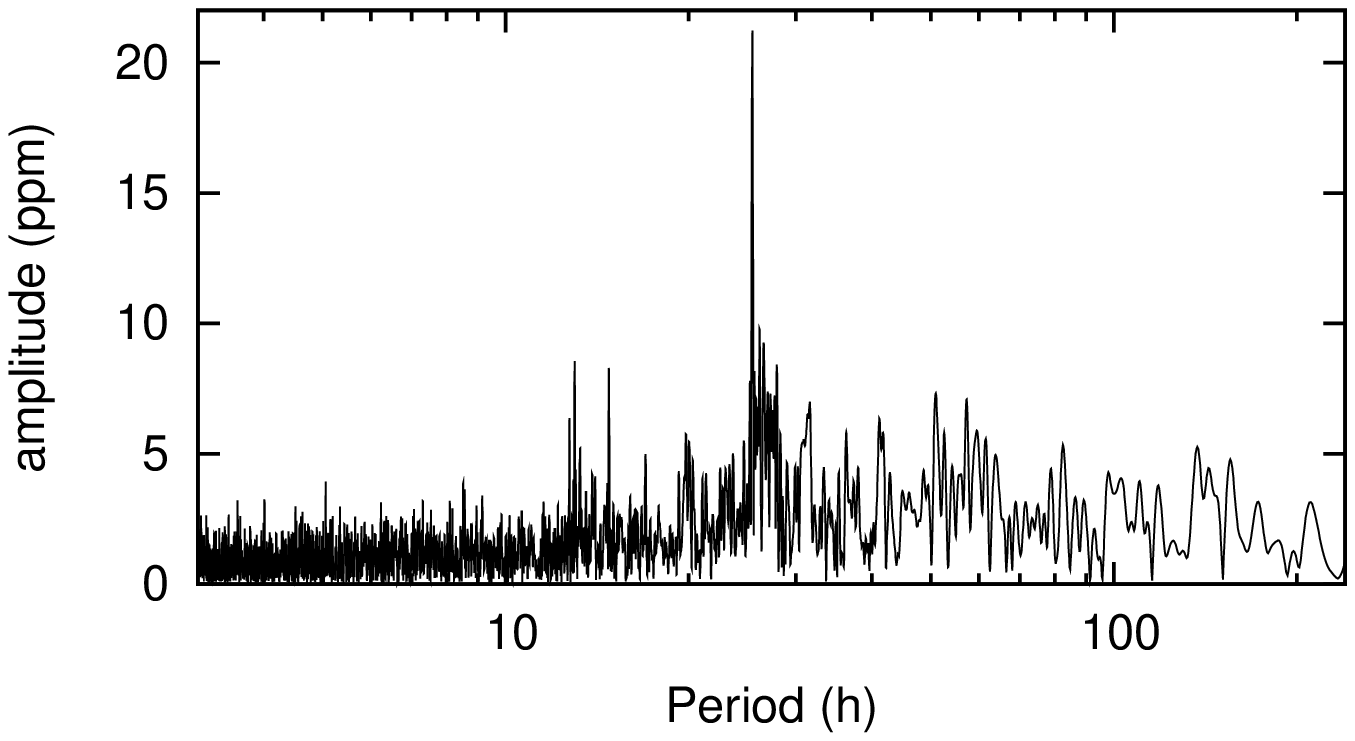}
\includegraphics[bb=111 239 454 509,width=8.1cm]{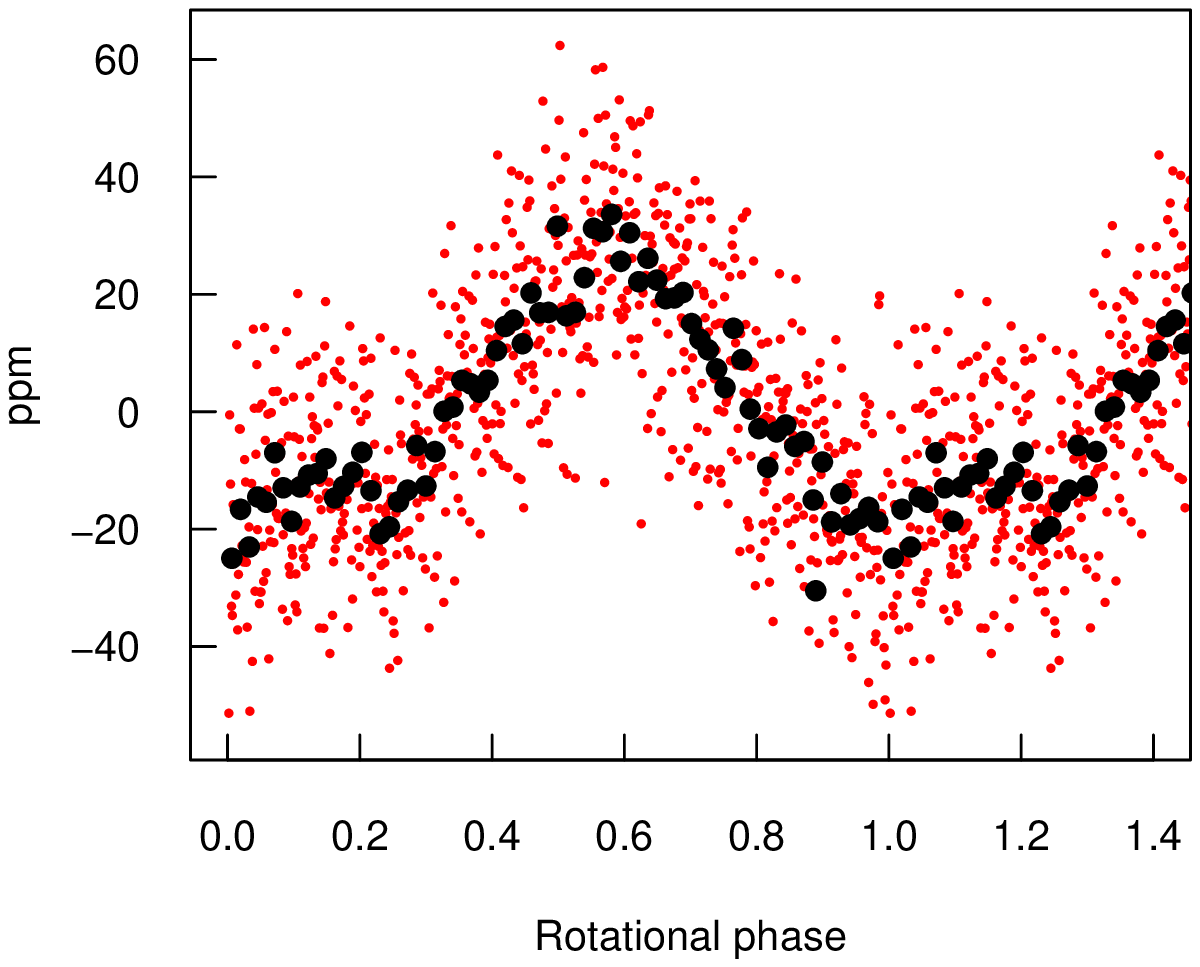}
\caption{Top: out-of-transit ligh tcurve of KOI-13, corrected for the third light. The reference flux level is KOI-13 A, i.e. the flux measured at eclipse. After subtracting this signal from the time series (middle upper panel), a signal with 25.432 hour period is detected with high significance (lower middle panel). Bottom: The phase diagram with 25.432 h period, after removing the orbit-dependent variations.}
\label{fig:fourplots}
\end{figure}

\begin{figure}
\includegraphics[bb=102 231 454 510,width=8.5cm]{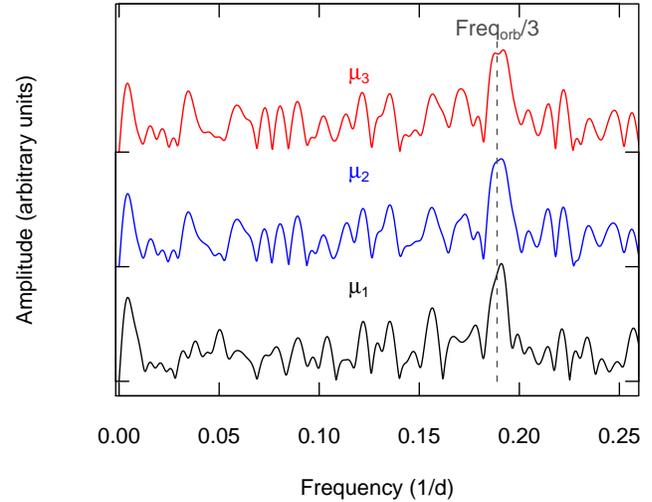}
\caption{Fourier-transforms of the light curve shape moments $\mu_1$--$\mu_3$ show modulations in the transit shape with a period of $3P_{\rm orb}$. The peaks around 0.189 1/day frequency ($=1/P_{orb}/3$) have $\approx$\,4-$\sigma$ confidence.}
\label{fig:periodograms}
\end{figure}

After removing the orbital phase-dependent light variations from the 
\emph{Kepler} photometry of KOI-13, we detected an additional periodic signal
that we argue below is due to rotation of the host star. The light curve was corrected 
with a dilution factor of $1.818$ (Szab\'o et al. 2011). The folded light curve with $P_{\rm orb}$ is plotted in the upper 
panel of Fig.~\ref{fig:fourplots} and we see the ellipsoidal and beaming variations (Mazeh et al. 2011, Shporer et al. 2011). The reference flux value is taken to be the centre of the 
secondary eclipse, when only KOI-13 A is visible. We smoothed 
and subtracted this variation from the out-of-transit light curve, and 
de-projected the residuals into the time domain (upper middle panel).
A prominent frequency was detected in these residuals at 0.9437 1/day, 
i.e. at $25.43\pm0.05$ hour period (Fig.~\ref{fig:fourplots}, 
lower middle panel). The semi amplitude of this variation is 21\,ppm, well 
above the average noise level of 1\,ppm in the 
Fourier spectrum. Harmonics of this frequency can also be detected \hl{up to the fourth order. 
This period has been found by Shporer et al. 2011 and Mazeh et al. 2011 independently.}
The binned phase diagram of the residuals with $25.43$ hour period is 
plotted in the bottom panel of Fig.~\ref{fig:fourplots}.

\hl{Both Mazeh et al (2011) and Shporer et al (2011) suggested a probable pulsational origin. Contrary to their proposition, we interpret it as the rotational period of KOI-13 A. The arguments for a rotational origin are the following:}
\begin{itemize}
\item{} \hl{The period is perfectly compatible with the expected rotational rate. Barnes et al. 2011 predicted
a 22--22.5 hour rotational period for KOI-13 A, depending on the stellar mass. They did not estimate
the error, but it is easy to calculate that errors in $v\sin i$ and in the stellar inclination results in $\pm 3.9$ hours. This prediction is in good agreement with the rotational origin of the new period.}
\item{} \hl{Balona (2011) detected signs of stellar rotation in 20\%{} of all A--F stars in the Kepler field caused by the granulation noise, and 8\%{} of them also exhibit starspot-like features in the light curves. Typically, these stars exhibit a dominant peak with 10--100 ppm amplitude at periods less than 3 days, the median is around 1 day for stars in the range of 7500--10,000 K. Another diagnostic is the scatter level at frequencies below $<50$/day, which exponentially increases by a factor of $\approx$1.6 toward low frequencies/high periods. The periodogram of KOI-13 looks exactly as described by Balona (2011), and the folded light curve (Fig.~1, lower panel) is also compatible with a rotating A-type star with magnetic activity.}

\item{} \hl{Stars similar to KOI-13 A can exhibit $p$ or $g$ mode pulsations ($\delta$ Sct and $\gamma$ Dor) or both (Uytterhoeven et al. 2011). However, the observed pulsations have periods less than one day, and several modes are observed with similar amplitudes. The general appearance of the frequency spectrum of KOI-13 does not resemble a pulsating star.}
\item{} \hl{Mazeh et al 2011 detected four harmonics of the 25.4-hour frequency in Q3 data, which we confirm here. The presence of such harmonics is typical for stellar rotation (Balona et al. 2011).}
\end{itemize}

\hl{It should be noted that both we and Mazeh et al. 2011 detected other frequencies between 1.5--2 c/day which are separated about equidistantly and have amplitudes of 4--7 ppm. Their harmonics do not appear in Q2+Q3 data, and these peaks do look compatible with pulsation. 

The source of the 25.4-hour period is the host star, KOI-13 A. This is seen from the systematic modulation of the transit shapes. Because this period is in 3:5 ratio with the orbital period,} every third transit occurs in front of the same stellar surface, and a modulation of the light curve shape is expected 
with a period of 3 transits. Since the individual light curves are too noisy for a direct comparison, we have 
to combine many data points and analyse their moments. Let us define $\mu_n$, the $n$-th light curve moment of each individual transit as
\begin{equation}
\mu_n :=  \sum\limits_{i\in\{\rm transit\}} \left( {t_i-C_i \over D}\right)^n\Delta f_i
\end{equation}
where $t_i$ and $f_i$ are the times and occulted fluxes belonging to 
each data points, $D$ is the transit duration and $C_i$ is the 
calculated mid-time of the transit, based on the ephemeris in Borucki 2011. Once 
the moments are assigned to the individual transits, a time series of 
moments can be analyzed in the standard fashion. In 
Fig.~\ref{fig:periodograms}, we plot the 
periodograms of the first three lightcurve moments (note that the 
amplitudes are in relative units). The periodograms show a detection with 5.27 day period, with a significance 
of $3-4$-$\sigma$ for each light curve moments, \hl{confirming that the 25.4-hour period modulates the shape of the transit with a period of $3 P_{orb}=5P_{rot}$.}

\subsection{The possible 5:3 spin--orbit resonance of KOI-13 A}

The 25.4-hour period is very close (within 0.1\%{}) to the 5:3 
spin-orbit resonance with KOI-13.01. \hl{Because of the large mutual inclination of 59$^\circ$ (Barnes et al. 2011),
the longitude of the companion varies with changing velocity.
Interestingly, the sinodical longitude of KOI-13.01 remains exactly constant (within 1\%{} fluctuations) on
$\approx1/8$ orbital arc surrounding the positions when the companion is at extremely high/low latitudes. Thus, KOI-13.01
and KOI-13 A move as if they were in exact 1:1 resonance for 3 hours.}

To date, theories of spin-orbit resonance cover the following cases: 
(i) spatial approximation with rigid bodies (e.g. Makarov 2011), where the body in resonance 
has little mass; (ii): resonant orbits of massive stars (e.g. Witte \&{} Savonije 2011); (iii) 
resonances in systems of compact bodies (e.g. Schnittman 2004). The case of resonance 
between the stellar spin and the orbit of a substellar companion is yet 
to be explored theoretically but, if confirmed, it will be a significant finding
in relation to the evolution of planetary systems.

\section{Transit duration variations}

\begin{figure}
\includegraphics[bb=30 80 431 666,width=8.3cm,clip]{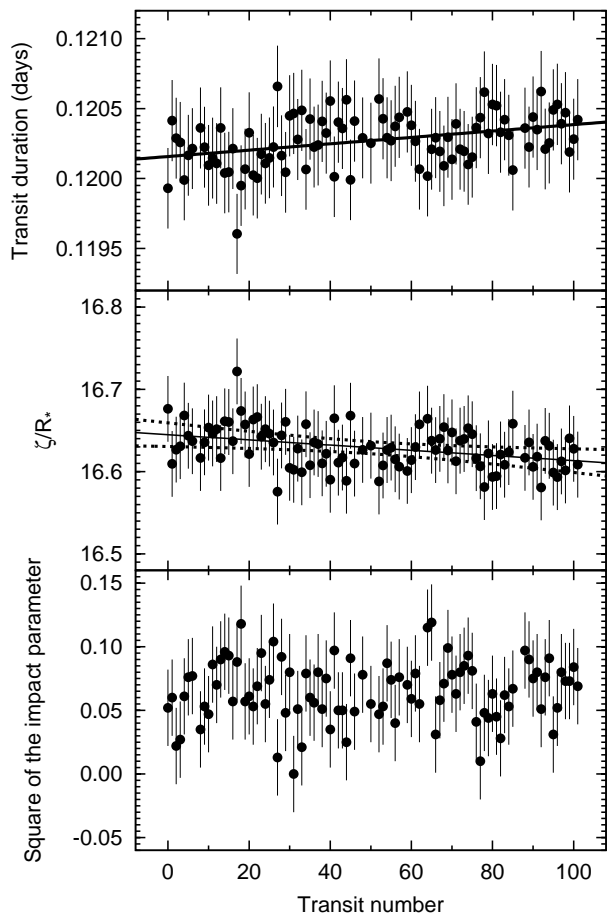} 
\caption{Variations in the transit duration (top panel), the reciprocal transit
duration (middle panel) and in the square of the impact parameter derived from transit shape 
($b^2$, lowest panel), as the function of the cycle number.
The plot of the reciprocal transit duration is superimposed by
the model of the best-fit linear trend together with the 3$\sigma$ errors of the linear regression.}
\label{fig:zetabeta}
\end{figure}

The analysis of the light curve moments suggested a long-period 
time-dependent variation in the transit curve shapes. 
By fitting each transits individually, 
we concluded that the duration of the transit is gradually increasing 
(Fig.~\ref{fig:zetabeta}, top panel). The significance of this finding is 3.85$\sigma$, 
with a false alarm probability of 0.012\%{}, based on an MCMC estimate. 
The 3-$\sigma$ confidence interval of the fitted linear regression is 
plotted in the middle panel of Fig.~\ref{fig:zetabeta}, 
onto the distribution of the 
reciprocal transit length (which is the actual output of the light 
curve fit algorithm). We now describe the details of how the light 
curves were analyzed.

First, every transit was fitted independently using the analytic models 
of \cite{mandel2002}. This fit invokes symmetric templates which cannot 
fit the known asymmetry of the light curves; but since the asymmetry 
appears similarly in each transits, we expect no time-dependent bias in 
the results. The free model parameters are: the transit time, the relative radius of the planet,
the reciprocal of the half duration, $\zeta/R_\star$, and the square of the impact 
parameter, $b^2$. These parameters are almost uncorrelated to each other (P\'al 2008). 
Indeed, our analysis shows that the inverse duration 
$\zeta/R_\star$ follows a secular trend (see Fig. 3).

The derived values of these parameters are displayed in 
Fig.~\ref{fig:zetabeta}. \hl{Since KOI-13 did not exhibit a transit timing variation, we can rule out short-term secular 
variations in the orbital semimajor axis of the transiting companion. 
Therefore, we conclude} that the decrease in 
$\zeta/R_\star$, indicating a lengthening of the transit duration, is due to 
the increasing orbital inclination (i.e. due to decreasing impact 
parameter). 
The observed linear trend in $\zeta/R_\star$ is ${\rm 
d}(\zeta/R_\star)/{\rm d}t=(-31.6\pm8.2)\cdot10^{-5}\,{\rm d}^{-1}\,{\rm 
cycle}^{-1}= (-17.9\pm4.6)\cdot10^{-5}\,{\rm d}^{-2}$. Since the 
relation between $\zeta/R_\star$, $a/R_\star$ and $b^2$ is
\begin{equation}
\left(\frac{a}{R_\star}\right)=\frac{\sqrt{1-b^2}}{n}\left(\frac{\zeta}{R_\star}\right),\label{eq:arzeta}
\end{equation}
\citep[see also][]{pal2008}, we can compute the time derivative of $b$ as
\begin{equation}
\dot b=\frac{{\rm d}b}{{\rm d}t}=\frac{1-b^2}{b}\left(\frac{\zeta}{R_\star}\right)^{-1}\frac{\rm d}{{\rm d}t}\left(\frac{\zeta}{R_\star}\right).
\end{equation}
This was obtained by re-ordering equation~(\ref{eq:arzeta}) and assuming that $a/R_\star$ 
is constant. By substituting our best-fit $b=0.253\pm 0.020$ into this 
equation, we had $\dot b=(-4.4\pm 1.2)\times 10^{-5}\,{\rm 
d}^{-1}=(-0.016\pm0.004)\,{\rm y}^{-1}$. \hl{This is a tiny change
in the transit duration, an order of magnitude smaller than the
precision with which $b$ can be determined from the light curve shape in the fitting 
procedure (Fig. 3, bottom panel). More precisely, the planet appeared to moved by only $\approx$15\%{} of its radius 
during the observed $\approx$100 transits, having no detectable
effect on the light curve shape. Consequently, we reasonably neglected the oblate shape of the rotating star
when converting transit duration to $b$. The stellar oblateness of a 2~M$_{\sun}$ star with radius 1.7~R$_{\sun}$
is 3\%{} (Murray \&{} Dermott, 1999), and less in an inclined projection. Therefore
the bias due to stellar oblateness is 1--2\%{} in the determined value of $\dot b$, 
much below the accuracy of its determined value. 
In Section 4, we will show that $\dot b$ is compatible with the secular perturbations caused by the oblateness of KOI-13~A, the rotating host star.
} 

\section{Interpretation}

\hl{
Assuming $M_p=9.2 \ M_J$ (see also Barnes et al. 2011), the angular momentum in the star and the companion are similar. Thus, the
axes of both the planet orbit and the stellar spin are precessing around the total angular momentum axis, opposing each other. To date, there is no exact theory for this case. In the observed examples, angular momentum of the orbit (double stars) or stellar spin (e.g. Mercury or lunisolar precession) dominates the other. One can study the precession of the orbital plane and the stellar spin under slightly different assumptions. We will show here that these lead to compatible results, and give a satisfactory estimate of the stellar oblateness.
}

\subsection{Secular J$_2$ perturbations}

As known from the 
theory of satellite motions \citep{kaula1966}, higher order moments of 
the gravitational potential of a host body yield periodic and secular 
perturbations in the orbits of nearby companions. The external 
gravitational potential of an extended body can be expressed as
\begin{equation}
V(r,\theta) = -{GM\over R} \left[ 1-\sum_{n=2}^\infty J_n \left( {R\over r} \right)^n \mathcal{P}_n (\cos\theta)\right],
\end{equation}
where $M$ is the total mass, $R$ is the equatorial radius, $J_n$ are 
constants and $\mathcal P_n$ are the Legendre polynomials. The  
most prominent perturbation is caused by $J_2$, due to the oblateness of 
the host body. MacCullagh's Theorem allows us compute $J_2$ using
\begin{equation}
J_2=\frac{1}{MR^2}\left(\Theta_{zz}-\frac{\Theta_{xx}+\Theta_{yy}}{2}\right) \approx {\Theta_{zz}-\Theta_{xx} \over MR^2},\label{eq:j2def}
\end{equation}
where $\Theta_{xx}=\Theta_{yy} \leq \Theta_{zz}$ are the principal 
moments of inertia. It is known that a non-zero $J_2$ results in secular 
perturbations in the angular orbital elements. Namely, the secular term in $\Omega$ (argument of 
ascending node) is computed as 
\begin{equation} 
\frac{{\rm d}\Omega}{{\rm d}t}=-\frac{3}{2}J_2n\left(\frac{a}{R}\right)^{-2}\frac{\cos \varphi}{(1-e^2)^2}.\label{eq:j2omega} 
\end{equation} 
Here $n$ denotes the orbital mean motion, $a$ is the semi-major axis and 
$e$ is the orbital eccentricity. 

It is known from vector geometry, that if
an unit vector $\mathbf{n}$ precesses around the unit vector $\mathbf{p}$ with
an angular frequency of $\omega_0$, the time derivative of $\mathbf{n}$
will be
\begin{equation}
\dot{\mathbf{n}}=\omega_0(\mathbf{p}\times\mathbf{n}).\label{eq:ncomponents}
\end{equation}
In our case, $\mathbf{n}$ is the unit vector parallel to the orbital
angular momentum of the transiting body.
Since the components of $\mathbf{n}$ are
\begin{equation}
\mathbf{n}=\left(
\begin{array}{c}
  \sin i\cos \Omega \\ 
  \sin i\sin \Omega \\ 
  \cos i \\
\end{array}\right)
\end{equation}
and the only observable quantity is $n_z\equiv\cos i$, we can write
\begin{equation}
\dot n_z = \omega_0 (p_xn_y-p_yn_x).
\end{equation}
By substituting $p_x=\sin i_{\rm p}\cos\Omega_{\rm p}$
and $p_y=\sin i_{\rm p}\sin\Omega_{\rm p}$ and the components $n_x$ and
$n_y$ from equation~(\ref{eq:ncomponents}), we obtain
\begin{equation}
\frac{{\rm d}\cos i}{{\rm d}t}=\omega_0\sin i\sin i_{\rm p}\sin\lambda,
\end{equation}
where $\Omega_p$ is the ascending node of the stellar equator, and $\lambda$ is the
longitude of the planet's ascending node, relative to that of the ascending node of the stellar equator
by definition.

\subsection{The inferred oblateness of the host star}

Assuming a circular orbit for the transiting companion and
substituting the above relation and equation~(\ref{eq:cosmutuali})
into equation~(\ref{eq:j2omega}), and by taking $\omega_0=d\Omega / dt$
as the precession rate inducted by $J_2$, we finally obtain
\begin{eqnarray}
\frac{{\rm d}\cos i}{{\rm d}t} & = & -\frac{3}{2}J_2 n\left(\frac{a}{R_*}\right)^{-2}\times \\
& & \times (\cos i\cos i_{\rm p}+ \sin i\sin i_{\rm p}\cos\Delta\Omega) \times\nonumber\\
& & \times \sin i\sin i_{\rm p}\sin\Delta\Omega.\nonumber
\end{eqnarray}
Since transits are observed, we can say here that $\cos i\ll\sin i\approx 1$.
In addition, $b=(a/R_\star)\cos i$, thus the above equation can be 
rearranged to give $\dot b$ as 
\begin{equation}
\dot b=-\frac{3}{2}J_2 n\left(\frac{a}{R_\star}\right)^{-1}\sin^2 i_{\rm p}\sin\lambda\cos\lambda.
\end{equation}
For $J_2$ we obtain
\begin{equation}
J_2\sin^2 i_{\rm p}\sin\lambda\cos\lambda = (3.8\pm 1.0)\cdot10^{-5}.
\end{equation}
Substituting the derived stellar parameters we find $J_2=(2.1\pm0.6)\times 10^{-4}$, and $d \Omega/dt = (3.4\pm0.9) \times 10^{-5}$/day.

\hl{Is this result compatible with a model star with the same mass, radius and rotational rate
as KOI-13 A? To check this, we calculated the moment of interia of $n=3$ polytrope models that were
distorted uniform stellar rotation.} The distortion of a rotating star is $(R-r)/R = \Omega^2 R^3 (2GM)^{-1} + 3 J_2/2$ \citep{stix2004}, where $R$ and $r$ are the radii at the equator and the poles, respectively. \hl{The $J_2$ term we derived for KOI-13 is two orders of magnitude less than the rotational term and can be neglected. The local distorions were calculated at at each internal point and then a $z$-contraction was applied. We assumed $dz/z=-\Omega^2 r^3 (2GM_r)^{-1}\equiv -3 \Omega^2 / (8 \pi G \langle \rho\rangle_{r})$, where $M_r$ and $\langle \rho\rangle_{r}$ are the mass and the mean density inside the radius $r$. This model is a first approximation that neglects higher-order distortions, but is enough precise for a rough estimate because the distortions are everywhere smaller than 3\%{}. The resulting values are $\Theta_{zz}=0.07760~MR^2$, $\Theta_{xx}=0.07743~MR^2$, $J_2=1.7\times10^{-4}$.}
Thus, the expected value of $J_2$ is in the range suggested by the theory of secular perturbations, supporting our interpretation.

\subsection{Precession rate}

\hl{Applying the framework developed for lunisolar precession, the precession rate of the star can be formulated as}
\begin{equation}
{d\Omega_p \over dt} = {3G M_p (\Theta_{zz}-\Theta_{xx}) \cos\varphi \over 2a^3 \Theta_{zz} n} = {3\over 2} J_2
{G M_p cos \varphi \over a^3 n \beta.}
\end{equation}
\hl{Here $M_p$ is the mass of the planet and $\beta=0.07760$ for KOI-13 A, as we derived in the previous section. Evaluating Eq 14, and applying the value of $J_2=2.1\times10^{-4}$ we find that $d\Omega_p/dt$ is in the range (3.81--1.67)$\times 10^{-5}$/day, for a companion of mass between 4--9.2~$M_J$. Therefore, $d\Omega_p/dt$ has the same order of magnitude that we derived for $d\Omega/dt$. The picture of a planet's plane precessing with $\approx500$ year period, and a star's orbital axis precessing with the same period, is self-consistent for KOI-13.}

\section{Summary}

With high-precision \emph{Kepler} photometry, we detected photometric and 
dynamical effects of stellar rotation in the exoplanet host KOI-13. 
From photometry alone, we have detected the following three phenomena for the first 
time:
\begin{itemize}

\item Stellar rotation in a probable resonance with the 
orbital period of the close-in substellar companion;

\item Long-period transit duration variations; and

\item Precession of the orbital plane of an exoplanet candidate.

\end{itemize}

Variations in transit duration were found from the fit of the consecutive 
light curves to transit shape models. Simultaneously, we fitted other 
model parameters such as the relative radius of the planet, the impact 
parameter and the timings of transits, and no variations were found in 
these quantities. The negative TTV detection is plausible because 
secular variations can be detected more easily in TDV than in transit timings 
for short time coverage \citep{pal2008k}. An argument against 
the reality of TDV can be that we fitted symmetrical templates to a 
transit with known slight asymmetry. However, we consider that the 
detected trend is not an artifact of this kind, because (i) there is no 
visible time-dependent trend in the asymmetric part of the light curves; 
(ii) the errors contain the ambiguity introduced by the asymmetric part, 
while the detection is still significant, and (iii) it is rather 
unlikely that the varying asymmetries affect only the transit duration, 
and leave the other parameters unchanged.

In interpreting the cause of transit duration variations, our hypothesis 
of secular $J_2$ perturbations is a probable scenario, but other 
processes cannot be excluded (e.g. perturbations from an unreported outer 
planet; the presence of a moon around KOI-13.01). However, the 
stellar rotation is the most plausible reason for TDV, because it does 
not invoke other bodies into the explanation, and the 
inferred $J_2$ is fully compatible with the expected value. In other 
words, one would predict secular $J_2$ perturbations of KOI-13.01 at 
the rate detected, taking into account the rapid rotation of the host 
star and the close proximity of its substellar companion.

\section*{Acknowledgments}

This project has been supported by the Hungarian OTKA Grants K76816, 
K83790 and MB08C 81013, the ``Lend\"ulet'' Program of the Hungarian 
Academy of Sciences, the ESA grant PECS~98073 and by the J\'anos Bolyai 
Research Scholarship of the Hungarian Academy of Sciences. Tim Bedding
is acknowledged for his comments.

{}

\label{lastpage}

\end{document}